\documentclass[sigplan,nonacm]{acmart}
\usepackage{soul}
\AtBeginDocument{%
  \providecommand\BibTeX{{%
    \normalfont B\kern-0.5em{\scshape i\kern-0.25em b}\kern-0.8em\TeX}}}

\setcopyright{acmcopyright}
\copyrightyear{2018}
\acmYear{2018}
\acmDOI{XXXXXXX.XXXXXXX}

\acmConference[Conference acronym 'XX]{Make sure to enter the correct
  conference title from your rights confirmation emai}{June 03--05,
  2018}{Woodstock, NY}
\acmPrice{15.00}
\acmISBN{978-1-4503-XXXX-X/18/06}

\begin{document}

\title[]{Emergent (In)Security of Multi-Cloud Environments}

\author{Morgan Reece}
\email{mlr687@msstate.edu}
\affiliation{%
  \institution{Mississippi State University }
  \country{USA}
}

\author{Theodore Lander Jr.}
\email{tel127@msstate.edu}
\affiliation{%
  \institution{Mississippi State University }
  \country{USA}
}

\author{Sudip Mittal}
\email{mittal@cse.msstate.edu}
\affiliation{%
  \institution{Mississippi State University }
  \country{USA}
}

\author{Nidhi Rastogi}
\email{nxrvse@rit.edu}
\affiliation{%
  \institution{Rochester Institute of Technology}
  \country{USA}
}

\author{Josiah Dykstra}
\email{josiah.dykstra@cyber.nsa.gov}
\affiliation{%
  \institution{National Security Agency}
  \country{USA}
}

\author{Andy Sampson}
\email{agsamps@uwe.nsa.gov}
\affiliation{%
  \institution{National Security Agency}
  \country{USA}
}

\renewcommand{\shortauthors}{Reece, et al.}

\begin{abstract}
As organizations increasingly use cloud services to host their IT infrastructure, there is a need to share data among these cloud hosted services and systems. A majority of IT organizations have workloads spread across different cloud service providers, growing their multi-cloud environments. When an organization grows their multi-cloud environment, the threat vectors and vulnerabilities for their cloud systems and services grow as well. The increase in the number of attack vectors creates a challenge of how to prioritize mitigations and countermeasures to best defend a multi-cloud environment against attacks. Utilizing multiple industry standard risk analysis tools, we conducted an analysis of multi-cloud threat vectors enabling calculation and prioritization for the identified mitigations and countermeasures. The prioritizations from the analysis showed that authentication and architecture are the highest risk areas of threat vectors. Armed with this data, IT managers are able to more appropriately budget cybersecurity expenditure to implement the most impactful mitigations and countermeasures.
\end{abstract}
\maketitle
\section{Introduction \& Background}



The proliferation of cloud-hosted applications continues to increase. While traditional cloud-hosted applications were designed to run in a \textit{single-cloud service}, nowadays more applications are being deployed to different cloud service providers (CSPs). For example, an organization may run its human resources application in Amazon Web Services (AWS) and run its IT Service Management in Microsoft Azure. The use of services that are hosted by different CSPs creates a \textit{multi-cloud environment} \cite{reece2023systemic}. Multi-cloud environments may be interconnected, enabling interoperability and data sharing.


The interoperability and data sharing between applications increase the attack surface and number of threats, including credential stealing, privilege elevation, and man-in-the-middle attacks. Ensuring security against these attack vectors targeting multi-cloud environments begins with risk assessment and vulnerability analysis. The risk assessment and vulnerability analysis can be divided into two categories: quantitative and qualitative. In our research, we used the STRIDE~\cite{hernan2006threat} and DREAD~\cite{howard2003writing} frameworks to conduct both qualitative and quantitative analysis.

\begin{figure} [ht]
    \centering
    \vspace{-4mm}
    \includegraphics[width=\linewidth]{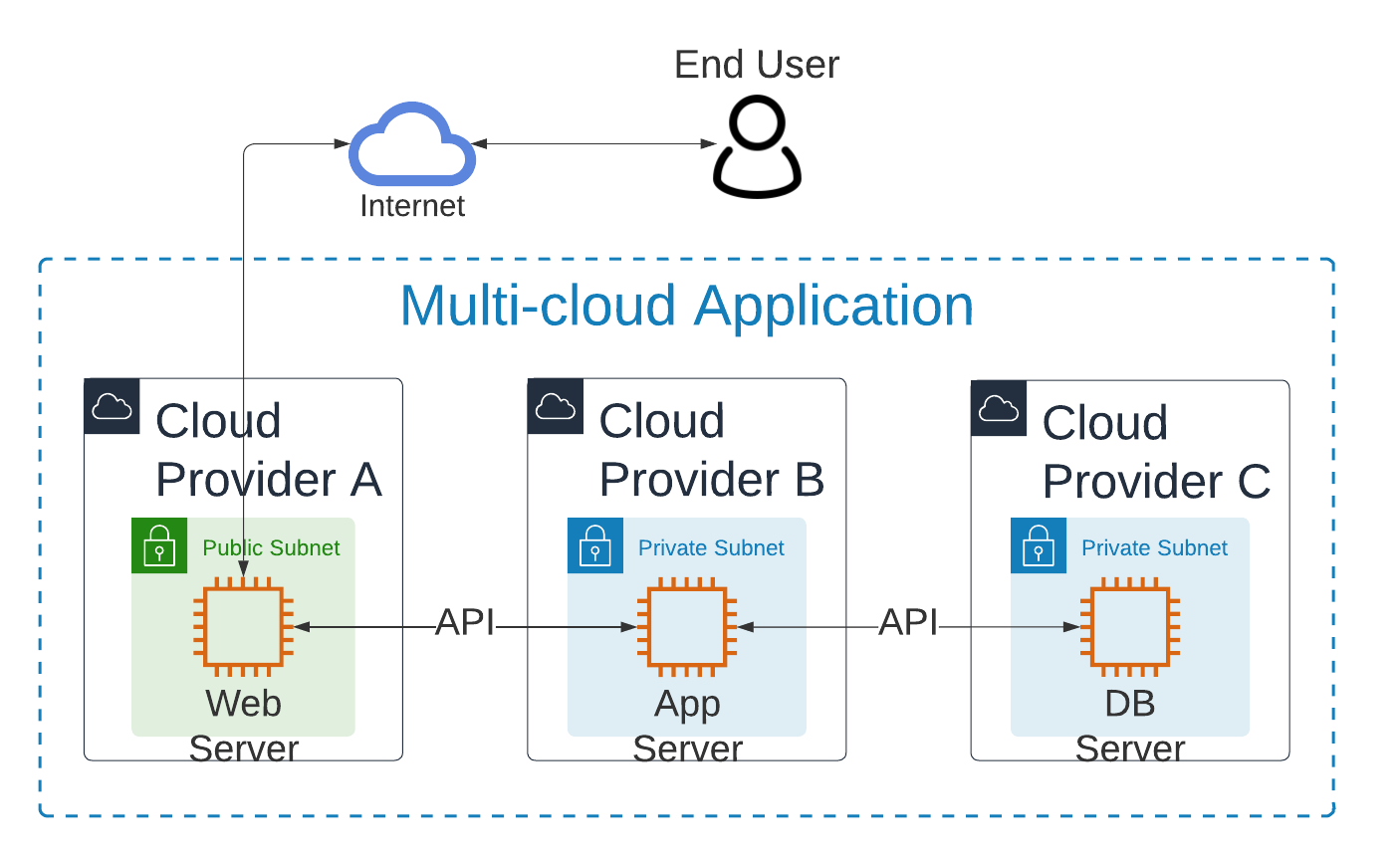}    
    \caption[A typical multi-cloud architecture and application.]{\textbf{Three-Tier Web Application Architecture.}}
    \label{fig:Multi-Cloud Blueprint}
    \vspace{-3mm}
\end{figure}

The objective of the analysis was to determine the breadth and scope of risks associated with multi-cloud attack vectors and the severity of each risk, allowing for the prioritization of the countermeasures and mitigations to prevent the associated threat~\cite{reece2023systemic}. To simulate and test the attacks and mitigations, we designed a generic architecture that emulates a multi-cloud environment (Figure~\ref{fig:Multi-Cloud Blueprint}). 
The architecture exposes the inter-process communication emulating the application programming interfaces (APIs) and interoperability in a hypothetical organization's multi-cloud environment. The simulation environment is a three-tier architecture where each tier (web server, app server, and database (DB) server) is deployed to a unique cloud provider.


As a realistic exemplar for the simulations and analysis, we used a generic healthcare provider as the business and technology model upon which to perform the research. Healthcare data is among the most valuable data to cyber criminals and must be well protected.

\section{Risk Analysis}\label{risk}
In the analysis, we applied two vulnerability and risk analysis models, STRIDE and DREAD, to the multi-cloud environment. The risk and vulnerability analysis began with qualitative risk analysis followed by quantitative risk analysis.

\subsection{Qualitative Risk Analysis}\label{qualitative}
The qualitative risk analysis aimed to identify and categorize the attack vectors of our defined multi-cloud environment. Utilizing the MITRE ATT\&CK framework~\cite{ATTACK} as guidance, we identified attack vectors focusing on methods that were specific to multi-cloud environments or those whose risk was exacerbated by a multi-cloud deployment. Once identified, we utilized the STRIDE attack methodology to categorize these attack vectors. Categorizing the attack vectors supported our identification of the threats, demonstrating our coverage of all aspects of the STRIDE methodology. Looking to the MITRE ATT\&CK framework, we identified mitigations for each of the threats. Along with the technical mitigations found in the MITRE ATT\&CK framework, we identified administrative countermeasures that would contribute to the reduction in the risk associated with the threat vector. With the qualitative risk analysis of the threats completed, the next step was to calculate an associated risk score for each threat through quantitative risk analysis.

\subsection{Quantitative Risk Analysis}\label{quantitative}
Our quantitative risk analysis started with the utilization of the DREAD risk assessment methodology~\cite{howard2003writing}. Our risk score estimate follows Fossen et al.~\cite{PractRisk}, where total damage is an average of four different types: legal, reputation, productivity, and other. We chose only to use the first three damage categories and not include the `Other' category seeing limited impact on the overall score by this category. Adding the average of the damage scores to the scores for the four threat probabilities gives the `Total Risk Score'.


After supplying values for the damages and threat probabilities and determining the resultant `Total Risk Score,' we ranked each threat vector. To support our analysis, we compared the resultant `Total Risk Score' with the corresponding score from the Exploit Prediction Scoring System (EPSS)~\cite{EPSS}. The comparison showed a direct correlation between the DREAD risk score and EPSS score confirming our DREAD damage and probability scores. With the resultant DREAD risk scores, organizations like those in our use case could prioritize the associated mitigation for each threat vector identified during our qualitative risk analysis. Given the relative priority of the DREAD risk score and the associated mitigations and countermeasures for the threat vectors, we can draw conclusions regarding the risk associated with multi-cloud specific threat vectors.

\section{Results}
Through performing the quantitative and qualitative analysis described in Section~\ref{risk}, we identified applicable mitigation to reduce the risk of the threat vector. The mitigations from the MITRE ATT\&CK framework are exclusively technical in nature and were not able to address all the risks associated with the threat vectors. Administrative countermeasures were specified to address risk that could not be mitigated through technical measures. These countermeasures were found in the Information Technology Infrastructure Library (ITIL) Foundation~\cite{ITIL_IBM} which provides guidelines and practices for IT service management. Table~\ref{tab:mitigations}, categorizes each attack vector, its corresponding countermeasures, and MITRE ATT\&CK framework mitigation to quantify its methodology. Along with tactics and techniques to reduce the risk from the identified threat vectors, we were able to use the risk score for each threat vector to prioritize the mitigations. 

\begin{table}[h]
  \centering
\caption{Attack Vector Countermeasures and Mitigations}
  \label{tab:mitigations}
\resizebox{1\columnwidth}{!}{%
\begin{tabular}{lll}
\textbf{Description of Threat}                                  & \textbf{Countermeasures}                      & \textbf{MITRE ATT\&CK Mitigation}          \\ \hline
\underline{\textbf{Architecture}} \\
DoS attacks                  & WAF w/DDoS mitigation                                 & Filter network traffic                     \\
Differing   Encryption       & ITIL - Change \& Secrets Management & N/A                                        \\
CVEs                         & Patch Management - System Hardening                   & Patch                                      \\
VPN Infiltration             & ICAM-MFA, Network segmentation                        & Network segmentation, MFA                  \\
Guest/Host OS's              & Patch Management - System Hardening                   & User Acct Mgmt                             \\
Additional Cloud Providers   & ITIL - Change Management - CMDB                       & N/A                                        \\
\underline{\textbf{Authentication}} \\
Session   Hijacking          & TLS encryption of sessions \& MFA                     & MFA, delete persistent cookies             \\
Substitution   Attack        & Secure Block-cypher - timestamp                       & Audit, PAM, Cert Mgmt                      \\
Man-in-the-Middle            & Secrets Management - DNSsec                           & Static network config                      \\
Inconsistent   User ACL      & ICAM - SCIM/SAML                                      & ICAM                                       
\end{tabular}
}
\end{table}

\section{Conclusion \& Future Work}
The conclusions drawn from the experiment and analysis show that the threats and vulnerabilities in multi-cloud environments are not necessarily unique to multi-cloud. However, we determined through risk analysis that in multi-cloud environments the priority of mitigation could be significantly different from a single-cloud environment. Our results suggest that a critical area for future research is securing identity in multi-cloud and understanding how identity can be presented as in a seamless manner across cloud providers.


\bibliographystyle{unsrt}
\bibliography{sample-base}

\end{document}